\documentclass{article}
\usepackage{authblk, amsfonts, amsmath, enumerate, blindtext, graphicx, amssymb, array, rotating}
\usepackage[a4paper, total={6in, 8in}]{geometry}
\usepackage[skip=1\baselineskip, indent=0 pt]{parskip}
\numberwithin{equation}{section}

\allowdisplaybreaks
\usepackage[bottom]{footmisc}

\title{Parametrizing Clifford Algebras' Matrix Generators with Euler Angles}
\author[1]{Manuel Beato Vásquez}
\author[1, 2]{Melvin Arias Polanco\thanks{melvin.arias@intec.edu.do}}
\affil[1]{Escuela de Física, Facultad de Ciencias, Universidad Autónoma de Santo Domingo, Av. Alma Mater, 10105, Dominican Republic}
\affil[2]{Laboratorio de Nanotecnología, Área de Ciencias Básicas y Ambientales, Instituto Tecnológico de Santo Domingo, Av. Los Próceres, Santo Domingo 10602, Dominican Republic}
\date{October 2023}

\begin{document}

\maketitle

\begin{abstract}

\noindent A parametrization, given by the Euler angles, of Hermitian matrix generators of even and odd-degenerate Clifford algebras is constructed by means of the Kronecker product of a parametrized version of Pauli matrices and by the identification of all possible anticommutation sets for a given algebra. The internal parametrization of the matrix generators allows a straightforward interpretation in terms of rotations, and in the absence of a similarity transformation can be reduced to the canonical representations by an appropriate choice of parameters. The parametric matrix generators of $2^{\mathrm{nd}}$ and $4^{\mathrm{th}}$-order are linearly decomposed in terms of Pauli, Dirac, and $4^{\mathrm{th}}$-order Gell-Mann matrices establishing a direct correspondence between the bases. In addition, and with the expectation for further applications in group theory, a linear decomposition of $GL(4)$ matrices on the basis of the parametric $4^{\mathrm{th}}$-order matrix generators and in terms of four-vector parameters is explored. By establishing unitary conditions, a parametrization of two sub-groups of $SU(4)$ is achieved.
\end{abstract}

\section{Introduction}

Historically, Pauli matrices and Dirac matrices allowed a significant breakthrough in the development of physics in the 20th century and even up to this date. The subsequent study of their properties reveals them as matrix representations of the generators of the Clifford algebras $Cl_{3,0}$ and $Cl_{1,3}$, or the so-called geometric algebra of physical space and geometric algebra of space-time, respectively \cite{1Hestenes, 2RukhsanUlHaq}. Among their many applications in physics \cite{3Gull-Lasenby-Doran, 4Traubenberg, 5-Baylis, 6Hildenbrand}, Clifford algebras play a fundamental role in the treatment of spinors in relativistic (and non-relativistic) quantum mechanics. Notably, Clifford algebras gain relevance in the standard model of particle physics by allowing the construction of spin groups \cite{7Wilsongroup, 8Wilsonproblem, 9baez}.

In the same fashion that Dirac's matrices enable the connection between the relativistic Hamiltonian (differential second-order) and the Dirac equation (differential first-order) \cite{10Dirac}, in general, matrix representations of Clifford algebras' generators serve as the connection between wave equations and first-order differential equations corresponding to a half-iteration. To clarify this point consider the two operators,

\begin{equation}
    \label{W}
    \hat{W}=-c^{-2}\partial_t^2+\sum_{\upsilon\,=\,1}^d\partial^2_\upsilon - \omega^2
\end{equation}

\begin{equation}
    \label{P}
    \hat{P} = ic^{-1}P_0\partial_t+\sum_{\upsilon\,=\,1}^dP_{\upsilon}\partial_{\upsilon}+i \omega P_{d+1}
\end{equation}

where $c, \omega \in \mathbb{R}$, the elements $P_\mu$ are square matrices, and the index `$d$' denotes the spatial dimensions. Operator (\ref{W}) yields a second-order differential equation with a structure of the Klein-Fock-Gordon equation, while operator (\ref{P}) yields a first-order differential equation with a structure of the Dirac equation. The condition on the matrices such that the action of the first-order operator on itself equals the second-order operator, $\hat{P}\hat{P}=\hat{W}$, is the anticonmmutator

\begin{equation}
\label{anticom}
    \{P_\mu, P_\nu\} = P_\mu P_\nu + P_\nu P_\mu = 2\delta_{\mu \nu}I, \quad \forall \mu, \nu 
\end{equation}

where $\delta_{\mu\nu}$ is the Kronecker delta and $I$ is the identity matrix (implicit in each term of $\hat{W}$). If the Dirac-like equation $\hat{P}\Psi=0$ is to be consistent with the continuity equation, then the matrices $P_\mu$ are to be Hermitian. That is,

\begin{equation}
\label{Herm}
    P_\mu^\dagger = (P_\mu^*)^T = P_\mu
\end{equation}

For our purposes, it suffices to say that the matrices satisfying eq. (\ref{anticom}) are the matrix representations of the generators of a Clifford algebra $Cl_n$ (in the particular case of the problem of reducing $2^{\mathrm{nd}}$-order operators to $1^{\mathrm{st}}$-order, $n=d+2$). The appearance of Pauli's and Dirac's matrices opened the way to the extensive study of such `matrix generators' (or `anticommuting involutions') and their properties from the 1920s and 1930s onward such that it is nowadays well established that \cite{11-Lee, 12-Pauli-german, 13-Pauli-f, 14-Newman, 15-Kestelman, 16-Hrubevs} $(n\in\mathbb{N})$:

\begin{enumerate}[(i)]
    \item The matrices that satisfy eq. ($\ref{anticom}$) are linearly independent, of even order, null trace, square determinant of unity, and eigenvalues of $\pm 1$.
    
    \item For even-order and degenerate odd-order Clifford algebras, $Cl_{2n}$ and $Cl_{2n+1}$, the minimum order of their matrix generators is $2^n$.
    
    \item The maximum amount of matrices of order $2^n$ that satisfy eq. ($\ref{anticom}$) is $2n+1$.
    
    \item If $P_\mu$ and $P'_\mu$  are two separate sets with an even number of matrices of order $2^n$ satisfying eq. ($\ref{anticom}$), then there exists a unique (up to a non-zero complex constant) and non-singular matrix $S$ such that
    
    \begin{equation}
    \label{similarity transf}
        P'_\mu = SP_\mu S^{-1}, \hspace{1cm}\mu=1, 2, ..., 2n    
    \end{equation}
    
    \item Let $U = (-1)^{\xi}P_1P_2\cdots P_{n-1}P_n$ be the product of all matrix generators of a Clifford algebra $Cl_{n}$ with a scaling factor such that $U^2=I$. The matrix $U$ anticommutes with each matrix generator if $n$ is even, and commutes if $n$ is odd.
\end{enumerate}

Conventional methodologies for constructing such matrix generators are typically based on the so-called \textit{canonical representations}, i.e. Pauli matrices as generators of $Cl_2$ \& $Cl_3$, and Dirac matrices as generators of $Cl_4$ \& $Cl_5$ \cite{19-Poole, 20-RCalvet, 21-Song-mat}. For higher order Clifford algebras, one takes the successive Kronecker product (also called direct or tensor product) between Pauli and Dirac matrices thus being able to obtain matrix generators in $Cl_6, Cl_7, Cl_8, Cl_9$ and still higher algebras by induction \cite{22-Gu, 23-Song-const, 24-Bilge}. For \textit{non-canonical representations}, their obtention and justification resort in property (iv), referred to in the literature as Pauli's (or Pauli-Dirac) Fundamental Theorem. This theorem is generalized for Clifford algebras of odd-order where the transformation between matrix generators, eq. (\ref{similarity transf}), adopts a slightly different structure since, in general, it incorporates the matrices $U$ of both sets \cite{17-Shikorov-extens, 18-Shirokov-calc}.

In his original 1927 paper where he introduces the now famous matrices that bear his name \cite{25-Pauli-zur, 26-Hill}, Wolfgang Pauli considers the following parametrized version of the spin matrices:

\begin{equation}
\label{R matrices}
    \begin{split}
    R_1 &=
    \begin{pmatrix}
    \sin\beta\,\sin\gamma                         & e^{-i\alpha}(\cos\gamma-i\cos\beta\,\sin\gamma) \\
    e^{i\alpha}(\cos\gamma+i\cos\beta\,\sin\gamma) & -\sin\beta\,\sin\gamma
    \end{pmatrix}\\
    R_2 &=
    \begin{pmatrix}
    \sin\beta\,\cos\gamma                             & -ie^{-i\alpha}(\cos\beta\,\cos\gamma-i\sin\gamma) \\
    ie^{i\alpha}(\cos\beta\,\cos\gamma + i\sin\gamma) & -\sin\beta\,\cos\gamma
    \end{pmatrix}\\
    R_3 &= 
    \begin{pmatrix}
    \cos\beta             & ie^{-i\alpha}\sin\beta \\
    -ie^{i\alpha}\sin\beta & -\cos\beta
    \end{pmatrix}\\
    \end{split}
\end{equation}

which results from a rotation through the Euler angles $\alpha, \beta, \gamma \in \mathbb{R}$ \cite{27-Arfken, 28-Goldstein} such that in vector notation,

\begin{equation}
\label{vector transf}
    \vec{R} = E \vec{\sigma}
\end{equation}

where $E = E_z(\gamma)E_x(\beta)E_z(\alpha)$ is the Euler rotation matrix (in `x-convention'). In matrix notation,

\begin{equation}
\resizebox{0.9\textwidth}{!}{
$\label{matrix transf}
    \begin{pmatrix}
        R_1\\
        R_2\\
        R_3
    \end{pmatrix}
    =
    \begin{pmatrix}
    \cos\alpha\cos\gamma - \sin\alpha\cos\beta\sin\gamma            & \sin\alpha\cos\gamma + \cos\alpha\cos\beta\sin\gamma            & \sin\beta\sin\gamma\\
    -\cos{\alpha}\sin{\gamma} - \sin{\alpha}\cos{\beta}\cos{\gamma} & -\sin{\alpha}\sin{\gamma} + \cos{\alpha}\cos{\beta}\cos{\gamma} & \sin{\beta}\cos{\gamma}\\
    \sin{\alpha}\sin{\beta}                                             & \cos{\alpha}\sin{\beta}                                             & \cos{\beta}
    \end{pmatrix}
    \begin{pmatrix}
    {\sigma}_1\\
    \sigma_2\\
    \sigma_3
    \end{pmatrix}$
    }
\end{equation}

Pauli was looking for consistency between transformations of the type (\ref{vector transf}) and `canonical transformations' given by eq. (\ref{similarity transf}) that respond to the spinorial wave-function transformation under a rotation of the coordinate system's axes, $\Psi'=S\Psi$, where $\Psi = (\Psi_1, \Psi_2)$. He proves that the coefficients of this transformation matrix $S$ are the complex conjugates of the Cayley-Klein parameters associated with the Euler angles of the rotation,

\begin{equation}
    S=
    \begin{pmatrix}
      \cos(\beta/2)\,e^{-i(\alpha+\gamma)/2} & -i\sin(\beta/2)\,e^{-i(\alpha-\gamma)/2}\\
      -i\sin(\beta/2)\,e^{i(\alpha-\gamma)/2} & \cos(\beta/2)\,e^{i(\alpha+\gamma)/2}
    \end{pmatrix}
\end{equation}

satisfying Hermiticity, $SS^{\dagger}=I$, and describing the transformation $R_{\mu}=S{\sigma}_{\mu}S^{\dagger},\, \mu=1, 2, 3$. Because it is always possible to diagonalize any one of the matrices into its `normal form', say $R_3$, then under the trivial rotation $\beta=2n\pi$ the matrices have a direct and convenient reduction:

\begin{equation}
    \begin{split}
        R_1(\beta=2n\pi) &=
        \begin{pmatrix}
            0 & e^{-i(\alpha+\gamma)} \\
            e^{i(\alpha+\gamma)} & 0
        \end{pmatrix}\\
        R_2(\beta=2n\pi) &=
        \begin{pmatrix}
            0 & -i\,e^{-i(\alpha+\gamma)} \\
            i\,e^{i(\alpha+\gamma)} & 0
        \end{pmatrix}\\
        R_3(\beta=2n\pi) &=
        \begin{pmatrix}
            1 & 0 \\
            0 & -1
        \end{pmatrix}
    \end{split}
\end{equation}

The phase $\alpha + \gamma$ now becomes arbitrary and upon an appropriate rotation of the z-axis can become zero such that the well-known Pauli $\sigma$-matrices are obtained.  It is verified that the set of matrices (\ref{R matrices}), just like the Pauli matrices, satisfy eqs. (\ref{anticom}) \& (\ref{Herm}). Hence, the set is a matrix representation of the generators of the Clifford algebra $Cl_3$.

In this paper we propose a construction of Hermitian, parametrized by Euler angles, matrix generators of even and odd-degenerate Clifford algebras through the Kronecker product of the parametrized Pauli matrices given by eq. (\ref{R matrices}). The methodology of construction, and specifically the identification of multiple sets of anticonmmutation in each algebra, shall be exposed in detail as it is not limited to the particular parametrization we take by choice\footnote{It is not even limited to the choice of the parameterized Pauli matrices \textit{per se}.}. The matrix representations we present are distinguished from the canonical by internally incorporating the parametrization of the Euler angles in the representation of the generators, allowing a direct interpretation of the matrices obtained with respect to the Pauli and Dirac matrices, and with the facility to revert to them without necessarily needing a similarity transformation but by an appropriate choice of parameters.

For the construction of the subsequent sets of matrices satisfying eqs. (\ref{anticom}) \& (\ref{Herm}), we will employ the properties and results (i) to (v). Notably, from (ii) and (iii) we will exploit the fact that by obtaining the matrix generators of a (degenerate) Clifford algebra of odd-order $Cl_{2n+1}$, we then immediately obtain the generators of the lower even-order Clifford algebra $Cl_{2n}$ by indiscriminately excluding any one of them. Conversely, from theorem (v) the matrix generators of an odd-degenerate algebra can be constructed from the $2n$ matrix generators of the even algebra $Cl_{2n}$ by including the matrix $U$ which anticommutes with each of them and thus constitutes itself as a generator.

Section \textbf{2} will serve as the starting point of the construction justifying the choice of the minimum-order matrices, $2^{\mathrm{nd}}$-order, starting from the restrictions on the elements that compose these matrices and identifying their connection with the matrices of the orthogonal group $O(3)$. By restricting to the $SO(3)$ group, the obtention of the Pauli matrices parametrized by the Euler angles is straightforward. In Section \textbf{3} we begin with the construction of matrix generators in higher-order algebras by constructing 16 parametrized matrices of $4^{\mathrm{th}}$-order via the Kronecker product of the previous matrices of lower order and identifying the multiple families of anticommutation that arise. In \textbf{3.1} we plan on characterizing their properties, and in \textbf{3.2} linearly decompose these matrices in terms of the Dirac and Gell-Mann matrices establishing a connection between bases of the matrix space of $4^{\mathrm{th}}$-order.  In section \textbf{4} we shall continue with the construction of parametrized matrices of $8^{\mathrm{th}}$ and $16^{\mathrm{th}}$-order, identifying the appropriate anticommutation sets. The induction pattern arises naturally from the various anticommutation structures of successive extensions, which will allow us to determine the categorizing structures of parametrized matrix generators of general odd- and even-order Clifford algebras, as well as the number of possible anticommutation sets if the generators of the algebra immediately below are given. Finally, in \textbf{5} we shall employ the parametric $4^{\mathrm{th}}$-order matrix basis to linearly decompose the general linear group of matrices $GL(4)$ and facilitate the construction of orthogonal and unitary sub-groups in terms of four-vector parameters.

\section{Construction of $2^{\mathrm{nd}}$-order Matrices}

Let us define a set of matrices of the lowest possible order, $2^{\mathrm{nd}}$, of constant and complex elements:

\begin{equation}
\label{dosuno}
    (R_\mu)_{mn}=\lambda_{mn}^\mu \in \mathbb{C}
\end{equation}

where $m,n=1,2$ and by (iii) of the previous section $\mu=1,2,3$. In order to satisfy eqs. (\ref{anticom}) \& (\ref{Herm}), the elements of such matrices are constrained by the following conditions:

\begin{align}
    &\lambda_{22}^\mu=-\lambda_{11}^\mu &&\textit{Null Trace} \\
    &(\lambda_{11}^\mu)^*=\lambda_{11}^\mu &&\textit{Hermiticity} \\
    &(\lambda_{12}^\mu)^*=\lambda_{21}^\mu &&\shortparallel \\
    &(\lambda_{11}^\mu)^2+|\lambda_{21}^\mu|^2 = 1 &&\textit{Involution} \\
    &2\,\lambda_{11}^\mu\,\lambda_{11}^\nu + \lambda_{21}^\mu\,(\lambda_{21}^\nu)^* +(\lambda_{21}^\mu)^*\,\lambda_{21}^\nu=0 \quad &&\textit{Anticommutation}
\end{align}

Eqs. (2.5) and (2.6) become significantly clearer with the notation 

\begin{equation}
    \lambda_{11}^\mu \equiv \lambda_3^\mu, \qquad \lambda_{21}^\mu \equiv \lambda_1^\mu + i \, \lambda_2^\mu, \qquad \lambda_j^\mu \in \mathbb{R}
\end{equation}

in such a way that the involution and anticommutation conditions condense into a single vector equation:

\begin{equation}
    \vec{\lambda}^{\mu}\cdot\vec{\lambda}^{\nu}=\delta^{\mu\nu}
\end{equation}

It is immediately seen that from eq. (2.8) we have six equations but nine parameters $\lambda_j^\mu$, thus yielding three free parameters. In the absence of an additional auxiliary condition, the $R_{\mu}$ matrices are inevitably parametrized. Moreover, eq. (2.8) is nothing more than the orthonormality condition for a set of vectors in the vector space of $\mathbb{R}_3$. Consequently, the nine parameters $\lambda_j^\mu$ of the $R_{\mu}$ matrices are conceived as the direction cosines between two orthonormal bases in $\mathbb{R}_3$. That is,

\begin{equation}
    \vec{\lambda}^{\mu}\cdot\vec{e}_j=\lambda_j^{\mu}
\end{equation}

From this realization we can associate the triad of orthonormal vectors $\vec{\lambda}^\mu$ as the constituents of the rows (or columns) of a $3^{\mathrm{rd}}$-order orthogonal matrix of elements

\begin{equation}
    (M)_{\mu j}=\lambda_j^{\mu}
\end{equation}

By imposing the additional restriction on $M$ as a matrix from the special orthogonal group, $M \in SO(3)$, then the interpretation of the parameters $\lambda_j^{\mu}$ as direction cosines admits Euler's rotation matrix as the choice of the direction cosines matrix. That is, the nine parameters of the $R_\mu$ matrices shall be given by the elements of the Euler rotation matrix, $(E)_{\mu j}=\lambda_j^{\mu}$, whereby employing the `x-convention\footnote{For different representations of the Euler rotation matrix, it is always possible to determine the similarity transformation matrix between distinct representations of the $R_\mu$ matrices.}', $E = E_z(\gamma)E_x(\beta)E_z(\alpha)$ with $\alpha, \beta, \gamma \in \mathbb{R}$  the Euler angles:

\begin{equation}
\resizebox{0.9\textwidth}{!}{%
     $E=
     \begin{pmatrix}
        \lambda_1^1 & \lambda_2^1 & \lambda_3^1 \\
        \lambda_1^2 & \lambda_2^2 & \lambda_3^2 \\
        \lambda_1^3 & \lambda_2^3 & \lambda_3^3 \\
    \end{pmatrix}
    =
    \begin{pmatrix}
        \cos\alpha\,\cos\gamma - \sin\alpha\,\cos\beta\,\sin\gamma            & \sin\alpha\,\cos\gamma + \cos\alpha\,\cos\beta\,\sin\gamma            & \sin\beta\,\sin\gamma\\
        -\cos{\alpha}\,\sin{\gamma} - \sin{\alpha}\,\cos{\beta}\,\cos{\gamma} & -\sin{\alpha}\,\sin{\gamma} + \cos{\alpha}\,\cos{\beta}\,\cos{\gamma} & \sin{\beta}\,\cos{\gamma}\\
        \sin{\alpha}\,\sin{\beta}                                             & \cos{\alpha}\,\sin{\beta}                                             & \cos{\beta}
    \end{pmatrix}$
}
\end{equation}

Therefore, the $R_\mu$ matrices take the explicit Pauli-parametric form given by eq. (\ref{R matrices}). This set of matrices verifies the properties (implicit sum in $\kappa$ and with $\varepsilon_{\mu \nu \kappa}$ the Levi-Civita symbol):

\begin{align}
    &\{R_\mu, R_\nu\}=2\,\delta_{\mu \nu}\,I \\
    &[R_\mu, R_\nu]=2\,i\,\varepsilon_{\mu \nu \kappa}R_\kappa \\
    &R_\mu^\dagger=R_\mu=R_\mu^{-1} \\
    &det(R_\mu)=-1 \\
    &Tr(R_\mu)=0 \\
    &R_\mu R_\nu = \delta_{\mu \nu}\,I+i\,\varepsilon_{\mu \nu \kappa}\,R_\kappa \\
    &\sum_{\mu\,=\,0}^3k_\mu\,R_\mu=0 \quad \Longleftrightarrow \quad k_\mu=0,\;\forall\mu \\
    &X=\sum_{\mu\,=\,0}^3\frac{1}{2}Tr(XR_\mu)R_\mu
\end{align}

Properties (2.12), (2.14), and (2.16) define the matrices. Property (2.17), which is obtained by adding the anticommutator (2.12) with the commutator (2.13) and simplifying, ensures the quasi-closure\footnote{Formally, the $R_\mu$ matrices do not form a subgroup of the unitary group $U(2)$ since they in and of themselves do not satisfy closure. The set of 16 matrices formed by $\pm R_\mu$ and $\pm\,iR_\mu$ do.} or cyclic product between the matrices of the set. By quasi-closure we mean that the product between any two matrices in the set yields a different matrix of the set, up to a factor of $\pm\,1$ and $\pm\,i$.  Defining the identity matrix of second-order as $R_0\equiv I$, property (2.18) represents the linear independence of the matrices. Finally, for $(X)_{mn}=x_{mn}$ a $2^{\mathrm{nd}}$-order matrix, property (2.19) establishes the $R_\mu$ matrices as generators of the space of second-order matrices. As a consequence of these last two properties, the $R_\mu$ matrices form a \textit{basis} in said space. The connection between the parametric and non-parametric Pauli matrices becomes apparent by using prop. (2.19) to determine the transformation coefficients and matrix that links both bases. Without much effort one finds that the elements of such a matrix are nothing more than the direction cosines $\lambda_j^\mu$ which make up the parametric matrices themselves:

\begin{equation}
    \lambda_j^\mu = \frac{1}{2}Tr(\sigma_jR_\mu)
\end{equation}

So the explicit transformation between the matrix bases is given by eqs. (\ref{vector transf}) and (\ref{matrix transf}). As expected, the parametric $R_\mu$ matrices are conceived as a rotation of the Pauli-vector defined by $\vec{\sigma}=(\sigma_1, \sigma_2, \sigma_3)$ through the Euler angles $\alpha, \beta, \gamma$. Under the transformation given by eq. (\ref{vector transf}), the properties (2.12) to (2.19) are preserved. Naturally, for a suitable choice of the Euler angles ($\beta=2n\pi, \alpha+\gamma=0$) the rotation matrix becomes unity. Or, in other words, the parametric $R_\mu$ matrices reduce to the Pauli $\sigma$ matrices. 

\section{Construction of $4^{\mathrm{th}}$-order Matrices}

The methodology to be followed in constructing higher-order matrix generators (corresponding to higher-order Clifford algebras) consists, for the time being, of taking the Kronecker product of the parametrized $R_\mu$ matrices which satisfy eqs. (\ref{anticom}) \& (\ref{Herm}) in second-order producing thereof fourth-order matrices. Let us define

\begin{equation}
    A_{\mu\nu} \equiv R_\mu \otimes R_\nu
\end{equation}

By the properties of the Kronecker product, and by including the identity matrix of second order $R_0$, it is verified that these $4^{\mathrm{th}}$-order matrices have the properties:

\begin{align}
    &A_{\mu\nu}^{-1}=A_{\mu\nu} \\
    &A_{\mu\nu}^\dagger = A_{\mu\nu} \\
    &det(A_{\mu\nu}) = 1 \\
    &Tr(A_{\mu\nu}) = 0, \qquad \forall\mu,\nu\; \text{non-zero simultaneously} \\ 
    &Tr(A_{00}) = 4
\end{align}

Eq. (3.1) defines 16 involutory, Hermitian matrices of determinant equal to unity and zero trace, except for the trace of the identity matrix of fourth-order $A_{00}$. With these properties assured, by (iii) we set out to find all possible sets of $2(n=2)+1=5$ anticommuting matrices. The following identities will be useful. 

\begin{align}
    &\{A_{\mu\nu}, A_{\rho\tau}\} = 2\,\delta_{\mu\rho}\,\delta_{\nu\tau}\,I-2\,\varepsilon_{\mu\rho\eta}\,\varepsilon_{\nu\tau\theta}\,A_{\eta\theta}, \qquad \text{no null index} \\
    &\{A_{\mu\nu}, A_{0\tau}\} = R_{\mu}\otimes\{R_\nu, R_\tau\} \\
    &\{A_{\mu\nu}, A_{\rho0}\} = \{R_\mu, R_\rho\}\otimes R_{\nu} \\
    &\{A_{\mu\nu}, A_{\mu\tau}\} = I\otimes\{R_\nu, R_\tau\} \\
    &\{A_{\mu\nu}, A_{\rho\nu}\} = \{R_\mu, R_\rho\}\otimes I
\end{align}

With $\mu=1$, by eq. (3.10) we identify the structure $A_{1\nu}=R_1\otimes R_\nu$ with $\nu=1,2,3$ as an anticommuting trio of matrices. Using eq. (3.7) in order to determine at least a fourth matrix that anticommutes with the trio requires that, independently of the value that takes $\nu$, the anticommutator vanishes for $\delta_{1\rho}$ and $\epsilon_{1\rho\eta}$ both zero simultaneously. This is not possible, so we conclude that a matrix $A_{\rho\tau}$ with non-null indices cannot anticommute with all three matrices of the set $A_{1\nu}$. The anticommutator of eq. (3.8) with $\mu=1$ cannot possibly vanish for all $\nu$ because its own internal anticommutation depends on the particular values $\nu$ takes; the same conclusion as above follows. So far we have exhausted the cases in which none of the indices of $A_{\rho\tau}$ are null and when the first is. It remains the case for which the second index is null, and to this end we make use of identity (3.9). Independently of the value of $\nu$, the anticommutator (3.9) with $\mu=1$ vanishes for $\rho=2,3$ in virtue of the anticommutation of the $R$ matrices. Therefore, the matrices $A_{20}$ and $A_{30}$ anticommute with the matrices $A_{1\nu}$ for all $\nu$. Because it is guaranteed by eq. (3.11) that these two new matrices anticommute with each other ($\mu=2, \rho=3, \nu=0$), we have then obtained the following quintet of involutory, Hermitian, and anticommuting matrices: 

\begin{equation}
    \begin{split}
        &A_{11}=R_1 \otimes R_1, \qquad A_{12}=R_1 \otimes R_2, \qquad A_{13}=R_1 \otimes R_3 \\
        &A_{20}=R_2 \otimes R_0, \qquad A_{30}=R_3 \otimes R_0
    \end{split}
\end{equation}

Following the same procedure we have described, for the trios $A_{2\nu}$ and $A_{3\nu}$ we find two distinct anticommutation quintets:

\begin{equation}
    \begin{split}
        &A_{21}=R_2 \otimes R_1, \qquad A_{22}=R_2 \otimes R_2, \qquad A_{23}=R_2 \otimes R_3 \\
        &A_{10}=R_1 \otimes R_0, \qquad A_{30}=R_3 \otimes R_0
    \end{split}
\end{equation}

\begin{equation}
    \begin{split}
        &A_{31}=R_3 \otimes R_1, \qquad A_{32}=R_3 \otimes R_2, \qquad A_{33}=R_3 \otimes R_3 \\
        &A_{10}=R_1 \otimes R_0, \qquad A_{20}=R_2 \otimes R_0
    \end{split}
\end{equation}

All 15 matrices (excluding the identity) participate in at least one of the three anticommutation quintets, with the exception of the $A_{0\tau}$ trio. Although these matrices anticommute with each other, from eq. (3.8) it is clearly visible that the anticommutator with a fourth matrix does not vanish independently of the value of $\tau$. As is evident, this result is true for all cases whether the indices of $A_{\mu\nu}$ are null or not. Thus, it is not possible to find a fourth matrix that anticommutes with the trio $A_{0\tau}$. Given this impossibility, we set out to find the matrices that anticommute with each \textit{pair} of matrices with structure $A_{0\tau}$. For instance, take the pair $A_{01}$ and $A_{02}$; It follows from eq. (3.8) that the only value that vanishes this anticommutator for the values $\tau=1,2$ \textit{simultaneously} is $\nu=3$. This result is true for all values of $\mu$, which implies that the trio $A_{\mu3}$ anticommutes with the duo $A_{01}$, $A_{02}$. The anticommutation of this trio is granted by eq. (3.11) for $\nu=3$ and $\mu\neq\rho$ non-nulls, so a fourth quintet of anticommuting matrices is:

\begin{equation}
    \begin{split}
        &A_{13}=R_1 \otimes R_3, \qquad A_{23}=R_2 \otimes R_3, \qquad A_{33}=R_3 \otimes R_3 \\
        &A_{01}=R_0 \otimes R_1, \qquad A_{02}=R_0 \otimes R_2
    \end{split}
\end{equation}

Following the same procedure for the remnants pairs $A_{01}$, $A_{03}$ and $A_{02}$, $A_{03}$ the last possible anticommuting quintets are found:

\begin{equation}
    \begin{split}
        &A_{12}=R_1 \otimes R_2, \qquad A_{22}=R_2 \otimes R_2, \qquad A_{32}=R_3 \otimes R_2 \\
        &A_{01}=R_0 \otimes R_1, \qquad A_{03}=R_0 \otimes R_3
    \end{split}
\end{equation}

\begin{equation}
    \begin{split}
        &A_{11}=R_1 \otimes R_1, \qquad A_{21}=R_2 \otimes R_1, \qquad A_{31}=R_3 \otimes R_1 \\
        &A_{02}=R_0 \otimes R_2, \qquad A_{03}=R_0 \otimes R_3
    \end{split}
\end{equation}

It can be observed that every matrix, with the exception of the identity matrix, is now a part of at least two anticommutation quintets. These six quintets can be classified according to the following two structures. 

\begin{equation}
    A_{a\mu}=R_a \otimes R_\mu, \qquad A_{b0}=R_b \otimes I_2, \qquad A_{c0}=R_c \otimes I_2
\end{equation}

\begin{equation}
    A_{\mu a}=R_\mu \otimes R_a, \qquad A_{0b}=I_2 \otimes R_b, \qquad A_{0c}=I_2 \otimes R_c
\end{equation}

The indices $a,b,c$ take the values 1, 2, 3 cyclically ($a\neq b\neq c$), and for a fixed $(a, b, c)$ the index $\mu$ runs from 1 to 3. 

\subsection{Characterisation}

The anticommutators, eqs. (3.7)-(3.11), and the involution (and as will be seen briefly, the product between any two matrices) occur through the product between the $R$ matrices associated with the first indices, and the product between the $R$ matrices associated with the corresponding second indices\footnote{In other words, there are no `mixed' products between first and second index matrices.}. From this, the matrices associated with the first indices allow a different parameterization than the second-indices matrices, which then makes the matrices $A_{\mu\nu}$ hexa-parametric. That is, $A_{\mu\nu}=A_{\mu\nu}(\alpha_1,\beta_1,\gamma_1,\alpha_2,\beta_2,\gamma_2)$ . The products between all possible matrix structures are as follows.

\begin{align}
    &A_{\mu\nu}\,A_{\rho\tau} = \delta_{\mu\rho}\,\delta_{\nu\tau}\,A_{00} + i\,\delta_{\mu\rho}\,\varepsilon_{\nu\tau\theta}\,A_{0\theta} + i\,\delta_{\nu\tau}\,\varepsilon_{\mu\rho\eta}\,A_{\eta0} - \varepsilon_{\mu\rho\eta}\,\varepsilon_{\nu\tau\theta}\,A_{\eta\theta} \\
    &A_{\mu\nu}\,A_{0\tau} = \delta_{\nu\tau}\,A_{\mu0} + i\,\varepsilon_{\nu\tau\theta}\,A_{\mu\theta} \\
    &A_{\mu\nu}\,A_{\rho0} = \delta_{\mu\rho}\,A_{0\nu} + i\,\varepsilon_{\mu\rho\eta}\,A_{\eta\nu} \\
    &A_{0\nu}\,A_{\rho\tau} = \delta_{\nu\tau}\,A_{\rho0} + i\,\varepsilon_{\nu\tau\theta}\,A_{\rho\theta} \\
    &A_{0\nu}\,A_{0\tau} = \delta_{\nu\tau}\,A_{00} + i\,\varepsilon_{\nu\tau\theta}\,A_{0\theta} \\
    &A_{0\nu}\,A_{\rho0} = A_{\rho\nu} \\
    &A_{\mu0}\,A_{\rho\tau} = \delta_{\mu\rho}\,A_{0\tau} + i\,\varepsilon_{\mu\rho\eta}\,A_{\eta\tau} \\
    &A_{\mu0}\,A_{0\tau} = A_{\mu\tau} \\
    &A_{\mu0}\,A_{\rho0} = \delta_{\mu\rho}\,A_{00} + i\,\varepsilon_{\mu\rho\eta}\,A_{\eta0}
\end{align}

In eqs. (3.20)-(3.28) no index is zero and we do not consider the trivial products $A_{00}A_{\mu\nu}=A_{\mu\nu}$, etc. Eqs. (3.20), (3.24), and (3.28) reproduce the involutory nature of the matrices in the cases where $\mu=\rho$ and $\nu=\tau$. In general, these nine identities allow us to infer the quasi-closure of the set of 16 matrices $A_{\mu\nu}$ and the construction of their Cayley table in appendix A.2. Through these results it is possible to prove that:

\begin{align}
        &\sum_{\mu,\,\nu\,=\,0}^3 h_{\mu\nu}\,A_{\mu\nu}=0 \quad \Longleftrightarrow \quad h_{\mu\nu}=0,\;\forall\mu,\nu \\
        &Y=\sum_{\mu,\,\nu\,=\,0}^3 \frac{1}{4}Tr(YA_{\mu\nu})A_{\mu\nu}
\end{align}

where the $h_{\mu\nu}$ are complex constants and $(Y)_{mn}=y_{mn}$ is a $4^{\mathrm{th}}$-order matrix. As expected, these matrices are linearly independent and act as generators of the $4^{\mathrm{th}}$-order matrix space. Consequently, the set of matrices $A_{\mu\nu}$ constitutes a basis. Moreover, it is possible to reproduce all 16 matrices by employing only one of the anticommutation quintets. Because of its connection (which we'll show in section 3.2) with the Dirac gamma matrices, the Cayley sub-table of the $A_{2\nu}, A_{10}, A_{30}$ quintet is presented in appendix A.1. In terms of only this quintet, the linear decomposition of any $4^{\mathrm{th}}$-order matrix can be written as:

\begin{equation}
    \begin{split}
        Y=\;&h_{00}A_{00} + \sum_{\nu,\,\tau} h_{\nu\tau}A_{2\nu}A_{2\tau} + h_{10}A_{10} + \sum_\nu h_{1\nu}A_{2\nu}A_{30} + h_{20}A_{10}A_{30} + \sum_\nu h_{2\nu}A_{2\nu}\\
        &+h_{30}A_{30} + \sum_\nu h_{3\nu}A_{10}A_{2\nu}
    \end{split}
\end{equation}

It is also possible to construct the corresponding Cayley sub-tables and linear decompositions for the other anticommutation quintets. 

\subsection{Transformations between Bases}

But of course, the fifteen Dirac gamma matrices (in any representation) together with the identity also form a basis in the space of $4^{\mathrm{th}}$-order matrices. By means of eq. (3.30) or eq. (3.31) it is possible to linearly decompose Dirac's matrices in terms of the base of parametric matrices $A_{\mu\nu}$, and vice versa. Equivalently, from the definition itself, eq. (3.1), we can express the parametric matrices $R_{\mu}$ in terms of the basis of Pauli's matrices by eq. (\ref{matrix transf}) and identify the resulting Kronecker products with the respective Dirac matrices. In Dirac's representation, the gamma matrices are constructed as follows.

\begin{equation}
    \begin{alignedat}{2}
        &I = \sigma_0\otimes\sigma_0  &&i\gamma_{\mu}\gamma_{\nu}=\varepsilon_{\mu\nu\kappa}\sigma_0\otimes\sigma_{\kappa} \\
        &\gamma_5 = \sigma_1\otimes\sigma_0 &&\gamma_{0}\gamma_{\mu}=\sigma_1\otimes\sigma_{\mu} \\
        &i\gamma_0\gamma_5 = -\sigma_2\otimes\sigma_0 \hspace{2cm} &&i\gamma_{\mu}=-\sigma_2\otimes\sigma_{\mu} \\
        &\gamma_0 = \sigma_3\otimes\sigma_0 &&\gamma_{\mu}\gamma_{5}=\sigma_3\otimes\sigma_{\mu}
    \end{alignedat}
\end{equation}

In eq. (3.32) we have incorporated the imaginary unit in some of the matrices such that all of them, in addition to possessing zero trace, are Hermitian and involutory. The linear decomposition of the $A_{\mu\nu}$ matrices in terms of the Dirac matrices then gives

\begin{align}
    &A_{\mu\nu}=(\lambda_1^{\mu}\,\gamma_0 - i\,\lambda_2^\mu\,I - \lambda_3^{\mu}\,\gamma_5)\vec{\lambda^\nu}\cdot\vec{\gamma} \\
    &A_{\mu0}=\lambda_1^{\mu}\,\gamma_5 - \lambda_2^\mu\,i\gamma_0\gamma_5 + \lambda_3^{\mu}\,\gamma_0 \\
    &A_{0\nu}=\lambda_1^{\nu}\,i\gamma_2\gamma_3 + \lambda_2^\nu\,i\gamma_3\gamma_1 + \lambda_3^{\nu}\,i\gamma_1\gamma_2
\end{align}

where no index is null and $\vec{\gamma}=(\gamma_1, \gamma_2, \gamma_3)$. With the conditions under which the parametric matrices $R_{\mu}$ reduce to the Pauli matrices, $\lambda_j^{\mu}=\delta_j^{\mu}$, eqs. (3.33)-(3.35) reduce to

\begin{equation}
    \begin{alignedat}{4}
        &A_{00}=I && A_{01}\to i\gamma_2\gamma_3 && A_{02}\to i\gamma_3\gamma_1 && A_{03}\to i\gamma_1\gamma_2 \\
        &A_{10} \to \gamma_5 && A_{11} \to \gamma_0\gamma_1 && A_{12} \to \gamma_0\gamma_2 && A_{13} \to \gamma_0\gamma_3 \\
        &A_{20} \to -i\gamma_0\gamma_5 \hspace{1cm} && A_{21} \to -i\gamma_1 \hspace{1cm} && A_{22} \to -i\gamma_2 \hspace{1cm} && A_{23} \to -i\gamma_3 \\
        &A_{30} \to \gamma_0 && A_{31} \to \gamma_1\gamma_5 && A_{32} \to \gamma_2\gamma_5 && A_{33} \to \gamma_3\gamma_5
    \end{alignedat}
\end{equation}

Another basis in the $4^{\mathrm{th}}$-order matrix space is the fifteen Gell-Mann matrices, which are also Hermitian and null-trace, but their involution is cubic rather than square (with the exception of $\Lambda_8$ and $\Lambda_{15}$). Although in the literature the canonical symbology to represent these matrices is by the Greek lowercase letter $\lambda$, to avoid confusion with the parameters $\lambda_j^\mu$ we shall use the Greek capital letter $\Lambda$ to denote the Gell-Mann matrices of fourth-order. That said, to determine the linear decomposition of the $A_{\mu\nu}$ matrices with respect to said basis it is quite useful to express the Gell-Mann matrices in terms of the Dirac matrices and then directly substitute the results already found (3.33)-(3.35). These two bases are related by the following equations.

\begin{equation}
    \begin{alignedat}{2}
        &\Lambda_1=\frac{1}{2}(i\gamma_2\gamma_3+\gamma_1\gamma_5) &&\Lambda_{13}=\frac{1}{2}(i\gamma_2\gamma_3-\gamma_1\gamma_5)\\
        &\Lambda_2=\frac{1}{2}(i\gamma_3\gamma_1+\gamma_2\gamma_5) &&\Lambda_{14}=\frac{1}{2}(i\gamma_3\gamma_1-\gamma_2\gamma_5)\\
        &\Lambda_3=\frac{1}{2}(i\gamma_1\gamma_2+\gamma_3\gamma_5) \\
        &\Lambda_4=\frac{1}{2}(\gamma_5+\gamma_0\gamma_3) &&\Lambda_{11}=\frac{1}{2}(\gamma_5-\gamma_0\gamma_3)\\
        &\Lambda_5=-\frac{1}{2}i(\gamma_0\gamma_5+\gamma_3) &&\Lambda_{12}=-\frac{1}{2}i(\gamma_0\gamma_5-\gamma_3)\\
        &\Lambda_6=\frac{1}{2}(\gamma_0\gamma_1-i\gamma_2) &&\Lambda_9=\frac{1}{2}(\gamma_0\gamma_1+i\gamma_2)\\
        &\Lambda_7=-\frac{1}{2}(i\gamma_1+\gamma_0\gamma_2) &&\Lambda_{10}=-\frac{1}{2}(i\gamma_1-\gamma_0\gamma_2)\\
        &\Lambda_8=\frac{1}{\sqrt{3}}\Big(\gamma_0+\frac{1}{2}\gamma_3\gamma_5-\frac{1}{2}i\gamma_1\gamma_2\Big) \qquad &&\Lambda_{15}=\frac{1}{\sqrt{6}}\Big(\gamma_0-\gamma_3\gamma_5+i\gamma_1\gamma_2\Big)
    \end{alignedat}
\end{equation}

The inversion of these relations for the matrices we are interested in is as follows.

\begin{equation}
    \begin{alignedat}{2}
        &i\gamma_1=-(\Lambda_7+\Lambda_{10}) &&i\gamma_2\gamma_3=\Lambda_1+\Lambda_{13} \\
        &i\gamma_2=\Lambda_9-\Lambda_6 &&i\gamma_3\gamma_1=\Lambda_2+\Lambda_{14} \\
        &i\gamma_3=-\Lambda_5+\Lambda_{12} &&i\gamma_1\gamma_2=\Lambda_3+\frac{1}{3}\Big(-\sqrt{3}\Lambda_8+\sqrt{6}\Lambda_{15}\Big) \\
        &\gamma_5=\Lambda_4+\Lambda_{11} &&i\gamma_0\gamma_5=-(\Lambda_5+\Lambda_{12})\\
        &\gamma_0=\frac{1}{3}\Big(2\sqrt{3}\Lambda_8+\sqrt{6}\Lambda_{15}\Big) \qquad &&
    \end{alignedat}
\end{equation}

Therefore, eqs. (3.33)-(3.35) in Gell-Mann basis are:

\begin{align}
    \begin{split}
    A_{\mu\nu}=&\Big[\frac{1}{3}\lambda_1^\mu\Big(2\sqrt{3}\Lambda_8+\sqrt{6}\Lambda_{15}\Big) - i\lambda_2^\mu I - \lambda_3^\mu(\Lambda_4+\Lambda_{11})\Big]\Big[i\lambda_1^\nu(\Lambda_7+\Lambda_{10}) \\
    &+i\lambda_2^\nu(\Lambda_6-\Lambda_9) + i\lambda_3^\nu(\Lambda_5-\Lambda_{12})\Big]
    \end{split}
    \\
    A_{\mu0}=&\lambda_1^{\mu}(\Lambda_4+\Lambda_{11}) + \lambda_2^{\mu}(\Lambda_5+\Lambda_{12}) + \frac{1}{3}\lambda_3^\mu\Big(2\sqrt{3}\Lambda_8+\sqrt{6}\Lambda_{15}\Big) \\
    A_{0\nu}=&\lambda_1^\nu(\Lambda_1+\Lambda_{13}) + \lambda_2^\nu(\Lambda_2+\Lambda_{14}) + \lambda_3^\nu\Big[\Lambda_3+\frac{1}{3}\Big(-\sqrt{3}\Lambda_8+\sqrt{6}\Lambda_{15}\Big)\Big]
\end{align}
    
It can be seen that, unlike with the gamma matrices, the one-to-one correspondence between the matrices of both bases is lost upon the reduction $\lambda_j^\mu=\delta_j^\mu$.

\section{Construction of Higher Order Matrices}

\subsection{$8^{\mathrm{th}}$-order Matrices}

In line with the construction of the 4th-order matrices, let us define

\begin{equation}
    B_{\mu\nu\theta}=R_{\mu}\otimes R_{\nu}\otimes R_{\theta}=A_{\mu\nu}\otimes R_{\theta}=R_{\mu}\otimes A_{\nu\theta}
\end{equation}

This structure defines $4^3=64$ matrices of order $2^3=8$. The second and third equalities hold due to the definition (3.1) and the associative product of the Kronecker product. The following properties are verified. 

\begin{align}
    &B_{\mu\nu\theta}^{-1}=B_{\mu\nu\theta} \\
    &B_{\mu\nu\theta}^\dagger = B_{\mu\nu\theta} \\
    &det(B_{\mu\nu\theta}) = 1 \\
    &Tr(B_{\mu\nu\theta}) = 0, \hspace{3cm} \forall\mu,\nu,\,\theta\neq0\; \text{simultaneously} \\ 
    &Tr(B_{000})=Tr(I_8) = 4
\end{align}
    
With the involution, Hermiticity, and null trace assured let us determine the anticommutation sets. By (iii) we look for sets of $2(n=3)+1=7$ anticommuting matrices. Consider the anticommutator between two matrices with structure $A_{\mu\nu}\otimes R_{\theta}$,

\begin{equation}
    \{B_{\mu\nu\theta}, B_{abc}\}=A_{\mu\nu}A_{ab}\otimes R_\theta R_c+A_{ab}A_{\mu\nu}\otimes R_c R_\theta
\end{equation}

To identify the anticommutation septets let us first consider the case where at least one of the third indices $\theta, c$ is null. Arbitrarily but equivalently, let $c=0$ such that

\begin{equation}
    \{B_{\mu\nu\theta}, B_{ab0}\}=\{A_{\mu\nu}, A_{ab}\}\otimes R_\theta
\end{equation}

On the contrary, if none of the third indices are null then we can employ eq. (2.17) regarding the cyclic product of the $R$ matrices. By exploiting the symmetry of the Kronecker delta and the asymmetry of the Levi-Civita, we get

\begin{equation}
    \{B_{\mu\nu\theta}, B_{abc}\}=\{A_{\mu\nu}, A_{ab}\}\otimes \delta_{\theta c}\,R_0 + [A_{\mu\nu}, A_{ab}]\otimes i\,\varepsilon_{\theta cd}\,R_d
\end{equation}

We know that the sub-anticommutator $\{A_{\mu\nu}, A_{ab}\}$ becomes zero for any of the six families (quintets) of anticommuting matrices of 4th-order given by eqs. (3.18) \& (3.19). With these choices, the general anticommutator (4.7) vanishes for $\theta=c$ in both cases whether these indices are null or not. Following the same argument for the equivalent expression $R_{\mu}\otimes A_{\nu\theta}$, the general anticommutator vanishes for $\mu=a$ and $A_{\nu\theta}, A_{bc}$ being two matrices of the same anticommutation family. Hereafter we reserve the indices $A_{\pi\varsigma}$ to refer exclusively to any one of the six anticommutation families given by eqs. (3.18) \& (3.19). With this notation, the structures $B_{\mu\pi\varsigma}$ and $B_{\pi\varsigma\theta}$ conform $2\cdot4\cdot6=48$ anticommuting quintets of 8th-order\footnote{For each of the structures, given a fixed family $\pi\varsigma$ we'll have four quintets corresponding to the choices $\mu,\theta=0,1,2,3$. And, for a fixed $\mu,\theta$ we'll have six families $\pi\varsigma$.}. It remains to determine the two matrices that will form the septet. 

Let $\pi\varsigma$ and $\pi'\varsigma'$ be two matrices of the same anticommuting family of 4th-order. For the matrices $B_{\pi\varsigma\theta}$ and $B_{\pi'\varsigma'0}$ eq. (4.8) does not produce a new matrix since (independently of $\mu$) this anticommutator is not zero for $\pi=\pi', \varsigma=\varsigma'$. But for $B_{\pi\varsigma\theta}$ and $B_{\pi'\varsigma'c}$ eq. (4.9) does not work either because if $\pi\neq\pi', \varsigma\neq\varsigma'$ then $\{A_{\pi\varsigma}, A_{\pi'\varsigma'}\}=0$ and the anticommutation requires $\varepsilon_{\theta c d}=0\, \therefore\, \theta = c$. On the other hand, if $\pi=\pi', \varsigma=\varsigma'$ then $[A_{\pi\varsigma}, A_{\pi'\varsigma'}]=0$ and the anticommutation requires $\delta_{\theta c}=0\,\therefore\,\theta\neq c$. Therefore, in this form we are not able to generate at least a sixth matrix that anticommutes simultaneously with the five matrices of $B_{\pi\varsigma\theta}$. The same argument applies to the structure $B_{\mu\pi\varsigma}$, so we conclude that the sixth and seventh matrices must have a different structure to that of $B_{\pi\varsigma\theta}$ and $B_{\mu\pi\varsigma}$. That is, they must not include any term $A_{\pi\varsigma}$. 

In general, the indices of the families $A_{\pi\varsigma}$ do not exclude being null, so the products $A_{\pi\varsigma} A_{ab}$ in the anticommutator (4.7) with $B_{\pi\varsigma\theta}$ and $B_{abc}$ will take many different forms depending on the indices $\pi\varsigma$. We aspire to factorize this equation in terms of some sub-anticommutator or sub-commutator just like in eqs. (4.8) and (4.9). Since it is not possible to have a sixth matrix that anticommutes with the matrices $A_{\pi\varsigma}$ but every matrix commutes with the identity, we then make the proposal $a=b=0$ such that eq. (4.7) reduces to

\begin{equation}
    \{B_{\pi\varsigma\theta}, B_{00c}\}=A_{\pi\varsigma}\otimes \{R_\theta,R_c\}
\end{equation}

Clearly, this anticommutator vanishes for $\theta\neq c$ with no null index by virtue of the anticommutation of the $R$ matrices. The analogous argument for $B_{\mu\pi\varsigma}$ is given by the anticommutator with $B_{a00}$. Ergo, for $a, b, c$ that takes the values $1,2,3$ cyclically we have the following two classifying structures:

\begin{equation}
    B_{a\pi\varsigma}=R_a \otimes A_{\pi\varsigma}, \qquad B_{b00}=R_b \otimes I_4, \qquad B_{c00}=R_c \otimes I_4
\end{equation}

\begin{equation}
    B_{\pi\varsigma a}=A_{\pi\varsigma} \otimes R_a, \qquad B_{00b}=I_4 \otimes R_b, \qquad B_{00c}=I_4 \otimes R_c
\end{equation}

where obviously $\{B_{b00}, B_{c00}\}=\{B_{00b}, B_{00c}\}=0$ for $b, c$ distinct and non-null. Eqs. (4.11) \& (4.12) define $2\cdot3\cdot6=36$ anticommuting septets of nona-parametric matrices.

\subsection{$16^{\mathrm{th}}$-order Matrices}

Defining,

\begin{equation}
    C_{abcd}=R_a\otimes R_b\otimes R_c \otimes R_d = A_{ab}\otimes A_{cd} = R_a\otimes B_{bcd} = B_{abc} \otimes R_d
\end{equation}

we have $4^4=256$ matrices of order $2^4=16$. As usual, these matrices possess similar properties to eqs. (3.2)-(3.6) and (4.2)-(4.6), namely involution, Hermiticity, null trace, unit determinant, and for this order $C_{0000}=I_{16}$. The identification of the sets of $2(n=4)+1=9$ anticommuting matrices is directly analogous to that of the matrices defined by (4.1). Let us reserve the indices $B_{xyz}$ to denote any one of the 36 septets of anticommutation given by eqs. (4.11) \& (4.12). From the anticommutators of structure $R_a\otimes B_{bcd}$, and with an analogous argument for structure $B_{abc} \otimes R_d$, we have

\begin{align}
    &\{C_{abcd}, C_{0fgh}\}=R_a\otimes\{B_{bcd}, B_{fgh}\} \\
    &\{C_{abcd}, C_{efgh}\}=\delta_{ae}R_0\otimes\{B_{bcd}, B_{fgh}\}+i\varepsilon_{aej}R_j\otimes[B_{bcd}, B_{fgh}]
\end{align}

For a fixed $a, h$ the structures $C_{axyz}$ and $C_{xyzh}$ are immediately identified as anticommuting septets. The eighth and ninth matrices are obtained  with the \textit{ansatz},

\begin{equation}
    \{C_{axyz}, C_{e000}\}=\{R_{a}, R_{e}\}\otimes B_{xyz}
\end{equation}

which is null for $a\neq e\neq 0$. Ergo,

\begin{equation}
    C_{axyz}=R_a \otimes B_{xyz}, \qquad C_{b000}=R_b \otimes I_8, \qquad C_{c000}=R_c \otimes I_8
\end{equation}

\begin{equation}
    C_{xyza}=B_{xyz} \otimes R_a, \qquad C_{000b}=I_8 \otimes R_b, \qquad C_{000c}=I_8 \otimes R_c
\end{equation}

are $2\cdot3\cdot36=216$ nonets of anticommuting, involutory, Hermitian, and dodeca-parametric matrices. For a fixed septet $B_{xyz}$, we have two structures wherein each one of them the indices $a, b, c$ run cyclically from one to three. 

\subsection{$2^{n}$-order Matrices}

From the anticommutation structures (3.18)-(3.19) of 4th-order, (4.11)-(4.12) of 8th-order, and (4.17)-(4.18) of 16th-order it is possible to induce the generalized structures for higher dimensional anticommuting matrices as well as the number of possible anticommutation sets. From hypothesis, assume we are given all families of $2n-1$ matrix generators $M_{pqr\cdots(n-1)}$ of order $2^{n-1}$ of an even and odd-degenerate Clifford algebras $Cl_{2(n-1)}$ and $Cl_{2n-1}$.  Then, the $2n+1$ matrix generators of the Clifford algebras $Cl_{2n}$ and $Cl_{2n+1}$ shall be constructed as the $n$-th Kronecker product of the $R$ matrices,

\begin{equation}
    R_\mu \otimes R_{\nu}\cdots\otimes R_{n} = R_\mu \otimes M_{\nu\cdots n}=M_{\mu\nu\cdots(n-1)}\otimes R_{n}
\end{equation}

Such that,

\begin{equation}
    R_a \otimes M_{pqr\cdots(n-1)}, \qquad R_b \otimes I_{2^{n-1}}, \qquad R_c \otimes I_{2^{n-1}}
\end{equation}

\begin{equation}
    M_{pqr\cdots(n-1)} \otimes R_a, \qquad I_{2^{n-1}} \otimes R_b, \qquad     I_{2^{n-1}} \otimes R_c
\end{equation}

are the two structures of $6^{n-1}$ sets of $2^{n}$-th order, $3n$-parametric, involutory, Hermitian, and anticommuting matrices. If only one of the anticommuting families is given, eqs. (4.20) \& (4.21) define only 6 sets. In Table 1 we make a summary of the results gathered so far.

\begin{table}[h]
\label{Table 1 Main Result}
\centering
\resizebox{\textwidth}{!}{%
\begin{tabular}[h]{|>{\centering}m{2cm} | >{\centering}m{1.5cm} | >{\centering}m{1.5cm} | >{\centering}m{1.5cm} | >{\centering}m{7.5cm}|}
\hline
 \textbf{Clifford Algebra} & \textbf{Matrix Order} & \textbf{Number of Para-meters} & \vspace{1.5ex} \textbf{Anticom-muting Sets} \vspace{1.5ex} & \textbf{Matrix Generators} \tabularnewline
 \hline
 \vspace{1.5ex} $Cl_2, Cl_3$ \vspace{1.5ex} & $2^1$ & $3$ & $1$ & $R_\mu = \big\{R_1,R_2,R_3$ \tabularnewline
 \hline
 $Cl_4, Cl_5$ & $2^2$ & $6$ & $6$ & \begin{align*}
     A_{\pi \varsigma} = 
     \begin{cases}
         \big\{ R_{a} \otimes R_{\mu} \quad R_b \otimes I_2 \quad R_c \otimes I_2\\
         \big\{ R_{\mu} \otimes R_{a} \quad I_2 \otimes R_b \quad I_2 \otimes R_c 
     \end{cases}
 \end{align*} \tabularnewline 
 \hline
 $Cl_6, Cl_7$           & $2^3$ & $9$   &  $6^2$    & \begin{align*}
     B_{xyz} = 
     \begin{cases}
         \big\{ R_{a} \otimes A_{\pi \varsigma} \quad R_b \otimes I_4 \quad R_c \otimes I_4\\
         \big\{ A_{\pi \varsigma} \otimes R_{a} \quad I_4 \otimes R_b \quad I_4 \otimes R_c 
     \end{cases}
 \end{align*} \tabularnewline
 \hline
 $Cl_8, Cl_9$ & $2^4$ & $12$ & $6^3$ & \begin{align*}
     C_{pqrs} = 
     \begin{cases}
         \big\{ R_{a} \otimes B_{xyz} \quad R_b \otimes I_8 \quad R_c \otimes I_8\\
         \big\{ B_{xyz} \otimes R_{a} \quad I_8 \otimes R_b \quad I_8 \otimes R_c 
     \end{cases}
 \end{align*} \tabularnewline
 \hline
 $Cl_{2n}, Cl_{2n+1}$ & $2^n$ & $3n$ & $6^{n-1}$ & \begin{align*}
     \begin{cases}
         \big\{ R_a \otimes M_{pqr\cdots(n-1)} \quad R_b \otimes I_{2^{n-1}} \quad R_c \otimes I_{2^{n-1}} \\
         \big\{ M_{pqr\cdots(n-1)} \otimes R_a \quad I_{2^{n-1}} \otimes R_b \quad I_{2^{n-1}} \otimes R_c
     \end{cases}
 \end{align*} \tabularnewline
 \hline
\end{tabular}}
\caption{Matrix Generators with associated Clifford Algebra}
\end{table}

\section{Linear Decompositions}

With the construction of matrix generators finished, we now investigate a possible application of the matrices found.

\subsection{Inverse Matrix \& Compositions}

Consider the linear decomposition of a $4^{\mathrm{th}}$-order matrix in the base of the $A_{\mu\nu}$ matrices. By explicitly substituting the $R_\mu$ matrices of first-index in eq. (3.30) we have,

\begin{equation}
    Y=\sum_{\nu\,=\,0}^3
    \begin{pmatrix}
        h_{0\nu}+\sum_{\mu}h_{\mu\nu}\lambda_3^\mu & \sum_\mu h_{\mu\nu}(\lambda_1^\mu-i\lambda_2^\mu) \\
        \sum_\mu h_{\mu\nu}(\lambda_1^\mu+i\lambda_2^\mu) & h_{0\nu}-\sum_{\mu}h_{\mu\nu}\lambda_3^\mu
    \end{pmatrix}
    \otimes R_\nu
\end{equation}

Let us define the parameters,

\begin{align}
    k_\nu &= h_{0\nu}+\sum_{\mu}h_{\mu\nu}\lambda_3^\mu \\
    n_0 &= \sum_\mu h_{\mu0}(\lambda_1^\mu-i\lambda_2^\mu) \\
    -n_\nu &= \sum_\mu h_{\mu\nu}(\lambda_1^\mu-i\lambda_2^\mu), \quad \nu>0 \\
    -l_\nu &= \sum_\mu h_{\mu\nu}(\lambda_1^\mu+i\lambda_2^\mu) \\
    m_0 &= h_{00}-\sum_{\mu}h_{\mu0}\lambda_3^\mu \\
    -m_\nu &= h_{0\nu}-\sum_{\mu}h_{\mu\nu}\lambda_3^\mu, \quad \nu>0
\end{align}

Such that, using vector notation, we are left with

\begin{equation}
    Y=
    \begin{pmatrix}
        k_0I + \vec{k}\cdot\vec{R} & n_0I - \vec{n}\cdot\vec{R} \\
        -l_0I - \vec{l}\cdot\vec{R} & m_0I - \vec{m}\cdot\vec{R}
    \end{pmatrix}
\end{equation}

By defining a second matrix with the same structure as eq. (5.8), $Y'=Y'(k', n', l', m')$, the coefficients of the matrix resulting of the composition $Y'' = Y'Y$ are as follows.

\begin{align}
    k_0''&=k_0'k_0+\vec{k'}\cdot\vec{k}-n_0'l_0+\vec{n'}\cdot\vec{l} \\
    \vec{k}''&=k_0'\,\vec{k}+\vec{k'}\,k_0+i\,\vec{k'}\times\vec{k}-n_0'\,\vec{l}+\vec{n'}\,l_0+i\,\vec{n'}\times\vec{l} \\
    n_0''&=k_0'n_0-\vec{k'}\cdot\vec{n}+n_0'm_0+\vec{n'}\cdot\vec{m} \\
    -\vec{n}''&=-k_0'\,\vec{n}+\vec{k'}\,n_0-i\,\vec{k'}\times\vec{n}-n_0'\,\vec{m}-\vec{n'}\,m_0+i\,\vec{n'}\times\vec{m} \\
    -l_0'' &= -l_0'k_0 - \vec{l'}\cdot\vec{k} - m_0'l_0 + \vec{m'}\cdot\vec{l} \\
    -\vec{l}'' &= -l_0'\,\vec{k} - \vec{l'}\,k_0 - i\,\vec{l'}\times\vec{k} - m_0'\,\vec{l} + \vec{m'}\,l_0 + i\,\vec{m'}\times\vec{l} \\
    m_0'' &= -l_0'n_0 + \vec{l'}\cdot\vec{n} + m_0'm_0 + \vec{m'}\cdot\vec{m} \\
    -\vec{m}'' &= l_0'\,\vec{n} - \vec{l'}\,n_0 + i\,\vec{l'}\times\vec{n} - m_0'\,\vec{m} - \vec{m'}\,m_0 + i\,\vec{m'}\times\vec{m}
\end{align}

This particular parametrization coincides with analogous proposals using Dirac matrices as a basis \cite{29-Red'kov, 30-Red'kov}. The determination of the coefficients of the matrix $Y'=Y'(k', n', l', m')$ such that $Y'$ is the inverse matrix of $Y$ (and hence $Y''=I$) is definitely not simple. However, by means of the transformation 

\begin{equation}
    \vec{v}=E\,\vec{\tilde{v}}
\end{equation}

we can transform basis via eq. (1.7),

\begin{equation}
    \vec{v}\cdot\vec{R}=\Big(E\,\vec{\tilde{v}}\Big)\cdot\Big(E\,\vec{\sigma}\Big)= \tilde{v}^T\,E^T\,E\,\sigma = \tilde{v}^T\,\sigma = \vec{\tilde{v}}\cdot\vec{\sigma}
\end{equation}

where we have exploited the orthogonality of the Euler matrix that allows for the transformation between bases, and where $\vec{v}$ is a vector that admits a $3\times1$ column matrix representation. With eq. (5.18), eq. (5.8) takes the structure:

\begin{equation}
    Y=
    \begin{pmatrix}
        k_0I + \vec{\tilde{k}}\cdot\vec{\sigma} & n_0I - \vec{\tilde{n}}\cdot\vec{\sigma} \\
        -l_0I - \vec{\tilde{l}}\cdot\vec{\sigma} & m_0I - \vec{\tilde{m}}\cdot\vec{\sigma}
    \end{pmatrix}
\end{equation}

With this structure\footnote{This structure arises from the linear decomposition in the Weyl-Dirac basis.}, we closely follow the procedure of Red'kov \textit{et alii} in \cite{29-Red'kov} to determine the coefficients of the inverse matrix $Y'=Y^{-1}=Y^{-1}(\tilde{k}^{-1},\tilde{n}^{-1},\tilde{l}^{-1},\tilde{m}^{-1};\sigma)$. Then, by applying the inverse transformation of eq. (5.18), $\vec{\tilde{v}}=E^T\,\vec{v}$, and simplifying we get:

\begin{align}
    k_0^{-1} = &\;|Y|^{-1}\Big[k_0(mm)+m_0(ln)+l_0(nm)-n_0(lm) + i\,(E^T\,\vec{l})\cdot(E^T\,\vec{m})\times(E^T\,\vec{n})\Big] \\
    \notag \\
    \begin{split}
       \vec{k}^{-1} = &\;|Y|^{-1}\Big\{-\vec{k}(mm) - \vec{m}(ln) - \vec{l}(nm) + \vec{n}(lm) + 2\,\vec{l}\times(\vec{n}\times\vec{m})\\
    & + i\,E\Big[ m_0(E^T\,\vec{n})\times(E^T\,\vec{l}) + l_0(E^T\,\vec{n})\times(E^T\,\vec{m}) + n_0(E^T\,\vec{l})\times(E^T\,\vec{m})\Big]\Big\}
    \end{split} \\
    \notag \\
    n_0^{-1} = &\;|Y|^{-1}\Big[-k_0(nm) + m_0(kn) - l_0(nn) - n_0(km) + i\,(E^T\,\vec{k})\cdot(E^T\,\vec{m})\times(E^T\,\vec{n})\Big] \\
    \notag \\
    \begin{split}
       \vec{n}^{-1} = &\;|Y|^{-1}\Big\{-\vec{k}(nm) + \vec{m}(kn) - \vec{l}(nn) - \vec{n}(km) + 2\,\vec{k}\times(\vec{m}\times\vec{n})\\
    & + i\,E\Big[ k_0(E^T\,\vec{m})\times(E^T\,\vec{n}) + m_0(E^T\,\vec{k})\times(E^T\,\vec{n}) + n_0(E^T\,\vec{m})\times(E^T\,\vec{k})\Big]\Big\}
    \end{split} \\
    \notag \\
    l_0^{-1} = &\;|Y|^{-1}\Big[k_0(ml) - m_0(kl) - l_0(km) - n_0(ll) + i\,(E^T\,\vec{m})\cdot(E^T\,\vec{l})\times(E^T\,\vec{k})\Big] \\
    \notag \\
    \begin{split}
       \vec{l}^{-1} = &\;|Y|^{-1}\Big\{\vec{k}(ml) - \vec{m}(kl) - \vec{l}(km) - \vec{n}(ll) + 2\,\vec{m}\times(\vec{k}\times\vec{l})\\
    & + i\,E\Big[ m_0(E^T\,\vec{l})\times(E^T\,\vec{k}) + k_0(E^T\,\vec{l})\times(E^T\,\vec{m}) + l_0(E^T\,\vec{m})\times(E^T\,\vec{k})\Big]\Big\}
    \end{split} \\
    \notag \\
     m_0^{-1} = &\;|Y|^{-1}\Big[k_0(ln) + m_0(kk) - l_0(kn) + n_0(lk) + i\,(E^T\,\vec{n})\cdot(E^T\,\vec{l})\times(E^T\,\vec{k})\Big] \\
     \notag \\
    \begin{split}
       \vec{m}^{-1} = &\;|Y|^{-1}\Big\{-\vec{k}(ln) - \vec{m}(kk) + \vec{l}(kn) - \vec{n}(kl) + 2\,\vec{n}\times(\vec{l}\times\vec{k})\\
    & + i\,E\Big[ n_0(E^T\,\vec{k})\times(E^T\,\vec{l}) + l_0(E^T\,\vec{k})\times(E^T\,\vec{n}) + k_0(E^T\,\vec{n})\times(E^T\,\vec{l})\Big]\Big\}
    \end{split}
\end{align}

where $(ab) = a_0b_0 - \vec{a}\cdot\vec{b}$ and $|Y| = det(Y)$.  Following as well the same procedure of \cite{29-Red'kov} to compute the determinant of the matrix $Y$, after the inverse transformation we get:

\begin{equation}
    \begin{split}
        |Y| =\;&(kk)(mm) + (ll)(nn) + 2(mk)(ln) + 2(lk)(nm) - 2(nk)(lm) + \\
        & + 2\,i\,\Big[ (E^T\vec{l})\cdot(E^T\vec{m})\times E^T(k_0\,\vec{n} + n_0\,\vec{k}) + (E^T\vec{k})\cdot(E^T\vec{n})\times E^T(m_0\,\vec{l} + l_0\,\vec{m}) \Big] \\
        & + 4(\vec{k}\cdot\vec{n})(\vec{m}\cdot\vec{l})-4(\vec{k}\cdot\vec{m})(\vec{n}\cdot\vec{l})
    \end{split}
\end{equation}

\subsection{Unitary Conditions}

With the inverse matrix structure determined, we may consider the conditions for which the matrix structure given by eq. (5.8) belongs to the special group of unitary matrices $SU(4)$. The imposition $Y^\dagger = Y^{-1}$ yields the following restrictions on the coefficients.

\begin{equation}
    \begin{alignedat}{8}
    & k_0^* = k_0^{-1} && \vec{k}^* = \vec{k}^{-1} && -l_0^* = n_0^{-1} && \vec{l}^* = \vec{n}^{-1} \\
    & n_0^* = -l_0^{-1} \qquad && \vec{n}^* = \vec{l}^{-1} \qquad && m_0^* = m_0^{-1} \qquad && \vec{m}^* = \vec{m}^{-1} \\
    & |Y| = 1
    \end{alignedat}
\end{equation}

This set of equations is not quite simple to solve. For such reason, consider the following special cases.

\subsubsection{$n_\nu = l_\nu = 0$}

Under this restriction eq. (5.8) takes the form

\begin{equation}
    Y=
    \begin{pmatrix}
        k_0I + \vec{k}\cdot\vec{R} & 0 \\
        0 & m_0I - \vec{m}\cdot\vec{R}
    \end{pmatrix}
\end{equation}

The composition reduces to

\begin{equation}
    \begin{split}
    k_0''&=k_0'k_0+\vec{k'}\cdot\vec{k} \\
    \vec{k}''&=k_0'\,\vec{k}+\vec{k'}\,k_0+i\,\vec{k'}\times\vec{k} \\
    m_0'' &= m_0'm_0 + \vec{m'}\cdot\vec{m} \\
    \vec{m}'' &= m_0'\,\vec{m} + \vec{m'}\,m_0 - i\,\vec{m'}\times\vec{m}
    \end{split}
\end{equation}

The determinant reduces to the simple expression

\begin{equation}
    |Y| = (kk)(mm)
\end{equation}

And the coefficients of the inverse matrix take the form

\begin{equation}
    \begin{split}
        k_0^{-1} &= (kk)^{-1}\,k_0 \\
        \vec{k}^{-1} &= -(kk)^{-1}\,\vec{k} \\
        m_0^{-1} &= (mm)^{-1}\,m_0 \\
        \vec{m}^{-1} &= -(mm)^{-1}\,\vec{m} \\
    \end{split}
\end{equation}

The unitary conditions then yield

\begin{equation}
    Y^\dagger=
    \begin{pmatrix}
        (kk)^{-1}(k_0I -\vec{k}\cdot\vec{R}) & 0 \\
        0 & (mm)^{-1}(m_0I + \vec{m}\cdot\vec{R})
    \end{pmatrix}
    , \quad (kk)(mm) = 1
\end{equation}

\subsubsection{$k_\nu = m_\nu = 0$}

Matrix form,

\begin{equation}
    Y=
    \begin{pmatrix}
        0 & n_0I - \vec{n}\cdot\vec{R} \\
        -(l_0I + \vec{l}\cdot\vec{R}) & 0
    \end{pmatrix}
\end{equation}

Composition and determinant,

\begin{equation}
\begin{split}
    k_0'' &= -n_0'l_0 + \vec{n'}\cdot\vec{l} \\
    \vec{k}'' &= -n_0'\vec{l} + \vec{n}'l_0 + i\,\vec{n'}\times\vec{l} \\
    m_0'' &= -l_0'n_0 + \vec{l'}\cdot\vec{n} \\
    \vec{m}'' &= l_0'\vec{n} - \vec{l}'n_0 + i\,\vec{l'}\times\vec{n} \\
    |Y| &= (ll)(nn)  
\end{split}
\end{equation}

Inverse matrix,

\begin{equation}
\begin{split}
    n_0^{-1} &= -(ll)^{-1}l_0 \\
    \vec{n}^{-1} &= -(ll)^{-1}\vec{l} \\
    l_0^{-1} &= -(nn)^{-1}n_0 \\
    \vec{l}^{-1} &= -(nn)^{-1}\vec{n}
\end{split}
\end{equation}

Such that,

\begin{equation}
    Y^\dagger=
    \begin{pmatrix}
        0 & -(ll)^{-1}(l_0I - \vec{l}\cdot\vec{R}) \\
        (nn)^{-1}(n_0I + \vec{n}\cdot\vec{R}) & 0
    \end{pmatrix}
    , \quad (ll)(nn) = 1
\end{equation}

Both of these cases can be considered as a parametrization of two subgroups in $SU(4)$. 

\section{Conclusions}

In synthesis, we have shown that if a set of three $2^{\mathrm{nd}}$-order Hermitian matrices are to be the matrix representation of the generators of a Clifford algebra $Cl_3$ then the components of their elements in their internal structure, $\lambda_j^\mu$, must be the direction cosines between two orthonormal bases in $\mathbb{R}_3$. By associating the trio of orthonormal vectors $\vec{\lambda^\mu}$ with the rows (or columns) of an orthogonal $3^{\mathrm{rd}}$-order matrix, we have been allowed to choose as the direction cosines matrix the Euler rotation matrix, where each matrix of the set is then parametrized by the three Euler angles and we explicitly obtained the parametrized Pauli matrices given by eq. (1.6). Although this choice is not unique it allows for a simple and direct interpretation: The vector (or equivalently, the space) formed by the parametrized $R$ matrices is the product of a rotation of the Pauli vector through the Euler angles, namely eq. (1.7).  Such a transformation in the `spin phase space' preserves the properties (2.12) to (2.19). But of course, this connection between the two matrix bases is no coincidence since the construction argument works the inverse route. Starting \textit{a priori} from the linear decomposition of a set of $2^{\mathrm{nd}}$-order matrices in terms of the Pauli basis and imposing the conditions (1.3) \& (1.4) we find that the coefficient associated with the identity matrix vanishes and that the other coefficients are nothing but the direction cosines between two orthonormal bases in $\mathbb{R}_3$, eq. (2.20). In other words, the transformation matrix between the two bases is an orthogonal direction cosines matrix, whereby imposing the additional restriction of making it an element of $SO(3)$ then the Euler rotation matrix is admitted. With the parametrization of the Euler angles and appropriate convention one promptly arrives at the parametric Pauli matrices. Under either perspective, a suitable choice of angles reduces the parametric matrices to the $\sigma$ matrices.

Subsequently, for the construction of Hermitian matrix generators of $4^{\mathrm{th}}$-order the Kronecker product between the parametric Pauli matrices and the identity matrix of $2^{\mathrm{nd}}$-order was used. Excluding the 4th-order identity, this procedure defines a base of fifteen involutory, Hermitian, hexa-parametric, and null trace matrices. With the help of convenient identities, we set out in detail the methodology to determine the six anticommuting quintets that can be classified according to the two structures of eq. (3.18) \& (3.19). \textit{I.e.}, any one of these six quintets of matrices generates the Clifford algebra $Cl_5$, although it is noted that any single quintet can reproduce the remaining five quintets (all remaining matrices of the basis, really) by a permutation of products between the matrices composing a given quintet. Moreover, the fifteen $A_{\mu\nu}$ matrices are linearly decomposed in terms of the Dirac matrix basis (in standard representation) through eqs. (3.33)-(3.35), and in terms of the Gell-Mann $4^{\mathrm{nd}}$-order matrix basis through eqs. (3.39)-(3.41). For specific values of the Euler parameters, $\lambda_j^\mu = \delta_j^\mu$, there is a direct correspondence between each element of the $A_{\mu\nu}$ matrices and the Dirac gamma matrices; this is not the case for the Gell-Mann matrices. Even though the interpretation of the connection between the parametric $R$ matrices and the Pauli matrices emerges naturally, the physical or geometrical interpretation of the $A_{\mu\nu}$ matrices with respect to the Dirac matrices remains to be determined.

In any case, the construction methodology consisting of defining Kronecker products of appropriate dimension and identifying the anticommuting sets is explicitly extended to matrices of higher orders for $Cl_7$ and $Cl_9$. From the evolution of the matrix generators and their structures in successive extensions, the inductive structure is then identified. Given all the anticommuting families of matrix generators of an even and odd-degenerate Clifford algebras $Cl_{2(n-1)}$ and $Cl_{2n-1}$, then the matrix generators of the Clifford algebras $Cl_{2n}$ and $Cl_{2n+1}$ are separately given by the two structures of eqs. (4.20) \& (4.21). These equations encapsulate the $6^{n-1}$ sets of $2^n$-th order, $3n$-parametric, involutory, Hermitian, and anticommuting matrix generators. If only one family is given, the equations yield only 6 separate sets. Table 1 makes a summary of the principal results obtained. It should be noted that throughout this paper we have intentionally made no distinction between the signatures of Clifford algebras for a given order, since from the matrix representation perspective the generators of $Cl_{p,\,q}$ are simply given by multiplying $q$ matrix generators of signature $Cl_{p+q,\,0}$ by the imaginary unit.

Finally, an application of the parametric 4th-order matrix basis was explored by means of the linear decomposition of the general linear group $GL(4)$ in terms of four-vector parameters. With the determination of the inverse matrix and determinant, establishing unitary conditions permitted a parametrization of two subgroups of $SU(4)$. The construction of further matrix groups and subgroups (pin, spin, etc.) with different conditions is also expected to be possible, if not desirable.

\bibliographystyle{ieeetr}
\bibliography{Bibliography}

\newpage

\section*{Appendix}

\subsection*{A.1 Cayley sub-table for the anticommuting quintet $A_{2\nu}, A_{10}, A_{30}$}

\begin{table}[h]
    \centering
    \begin{tabular}{|c|c|c|c|c|c|}
    \hline
        $\ast$   & $A_{10}$   & $A_{21}$   & $A_{22}$   & $A_{23}$   & $A_{30}$   \\ \hline
        $A_{10}$ & $A_{00}$   & $iA_{31}$  & $iA_{32}$  & $iA_{33}$  & $-iA_{20}$ \\ \hline
        $A_{21}$ & $-iA_{31}$ & $A_{00}$   & $iA_{03}$  & $-iA_{02}$ & $iA_{11}$  \\ \hline
        $A_{22}$ & $-iA_{32}$ & $-iA_{03}$ & $A_{00}$   & $iA_{01}$  & $iA_{12}$  \\ \hline
        $A_{23}$ & $-iA_{33}$ & $iA_{02}$  & $-iA_{01}$ & $A_{00}$   & $iA_{13}$  \\ \hline
        $A_{30}$ & $iA_{20}$  & $-iA_{11}$ & $-iA_{12}$ & $-iA_{13}$ & $A_{00}$   \\
    \hline
    \end{tabular}
    \label{A.1}
\end{table} 

\subsection*{A.2 Cayley table for the $A_{\mu\nu}$ matrices}

The main diagonal (highlighted) serves as an axis of symmetry, except for scale factors. Every matrix of the set (except the identity) anticommutes with eight distinct matrices and commutes with the remaining eight, including itself and the identity.

\includegraphics[width=\textwidth]{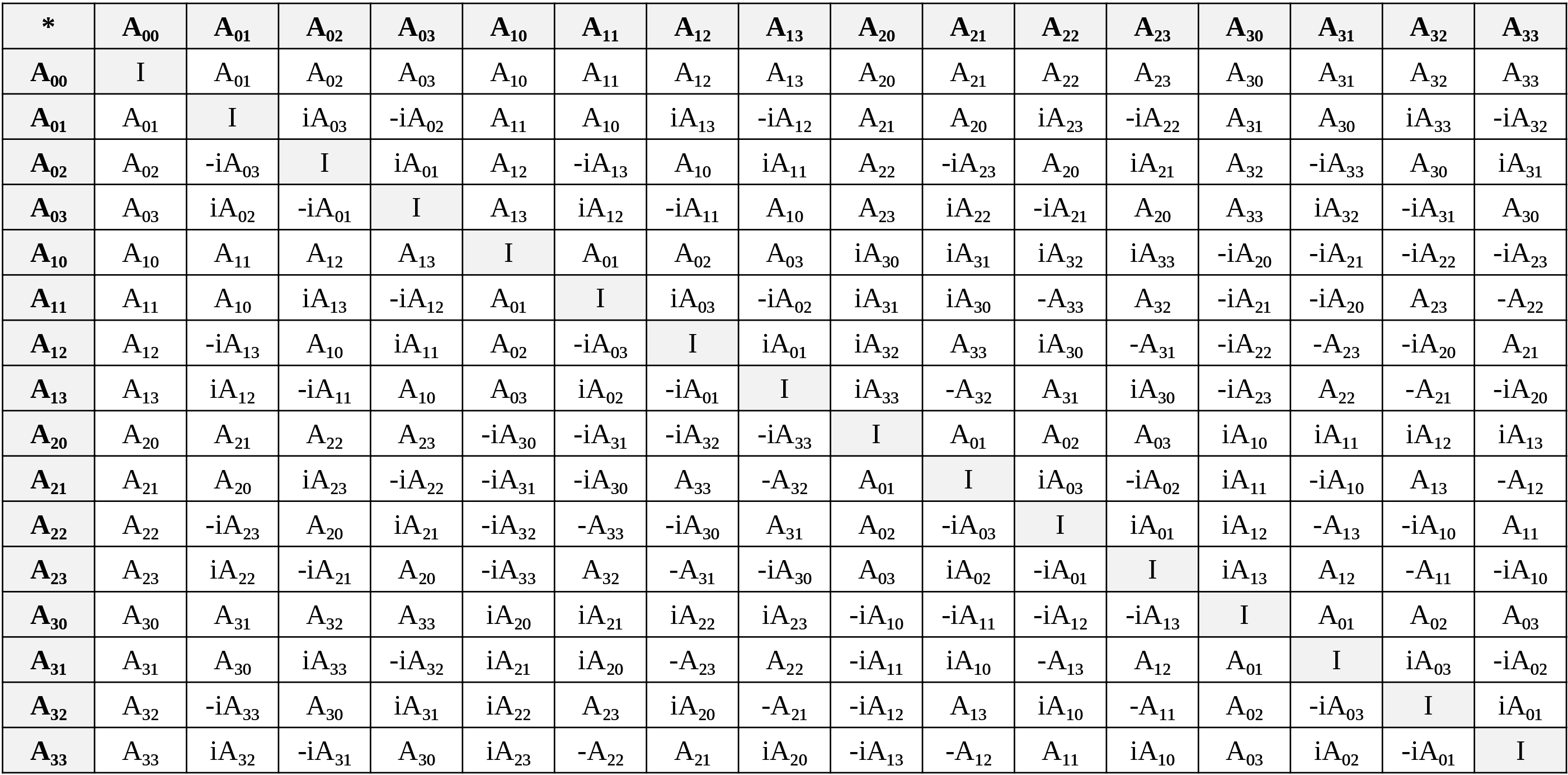}

\end{document}